# β-Ga$_2$O$_3$ Sub-Micron FinFETs with Si Delta-Doped Channel Modulating > 3x10$^{13}$ cm$^{-2}$ Charge Density

Nabasindhu Das, Cameron Gorsak, Ankita Kashyap, Advait Gilankar, Pushpanshu Tripathi, Salil A Paranjape, Trevor Thornton, Hari Nair, Nidhin Kurian Kalarickal

*Abstract*— This letter reports on the design and demonstration of high-performance β-Ga$_2$O$_3$ FinFETs utilizing MOCVD grown Si delta-doped channel to achieve enhanced carrier transport and electrostatic control. A record high sheet charge density of 3.3×10$^{13}$ cm$^{-2}$ was modulated using 100 nm fin channels, delivering a peak drain current of 410 mA/mm and a peak transconductance of 60 mS/mm. The FinFET architecture enables strong gate modulation achieving a high I$_{On}$/I$_{Off}$ ratio between 10$^8$-10$^9$. A low contact resistance of 0.42 Ω·mm was achieved to the Si delta doped channel using MOCVD contact regrowth. Small-signal RF characterization revealed a current gain cut-off frequency (f$_T$) of 3.8 GHz and maximum oscillation frequency (f$_{MAX}$) of 2.1 GHz for 0.8-micrometer gate length. These results demonstrate the efficacy of combining precision delta-doping and 3D FinFET geometry for high-frequency β-Ga$_2$O$_3$ electronics, establishing a platform for future RF and high-power applications.

## I. Introduction

Beta gallium oxide (β-Ga$_2$O$_3$) has gained significant attention as an ultra-wide bandgap semiconductor due to its large band gap (~4.8 eV), high breakdown field strength (~8 MV/cm), and favorable transport properties (μ=200 cm$^2$/V-s, v$_{sat}$=1.5-2 x 10$^7$ cm/s)[1], [2], [3]. These characteristics give β-Ga$_2$O$_3$ a Johnson's Figure of Merit (f$_T$V$_{BR}$=v$_{sat}$F$_{BR}$/2π) that is 2-3X higher than that of GaN, showcasing the promise of this material platform for RF and mm-wave electronics[4]. β-Ga$_2$O$_3$ RF transistors have shown rapid advancements over the past decade. Early MOSFETs achieved modest frequencies (f$_t$=3.3 GHz, f$_{max}$=12.9 GHz) using recessed gates and heavily doped contacts [5]. Significant improvement emerged with delta-doped channels, increasing f$_t$ to 27 GHz using 120 nm T-gates [6]. By using modulation-doped heterostructures, researchers demonstrated enhancement-mode HFETs reaching even higher frequencies (f$_t$=30 GHz, f$_{max}$=37 GHz), marking a critical breakthrough for β-Ga$_2$O$_3$ RF transistors [7].

This work is supported by the Army Research Office UWBG RF center under award No. W911NF2520005 and partly by the National Science Foundation under Grant No. NSF ECCS 2336397. Use of the core facilities at ASU is supported, in part, by NSF award ECCS-2025490.
Nabasindhu Das, Ankita Kashyap, Advait Gilankar, Trevor Thornton and Nidhin Kurian Kalarickal are with School of Electrical, Computer and Energy Engineering, Arizona State University, AZ, USA.
Cameron Gorsak, Pushpanshu Tripathi Salil A Paranjape and Hari Nair are with Department of Materials Science and Engineering, Cornell University, NY, USA.

More recently, thin-channel MOSFETs with optimized T-gates and surface passivation achieved remarkable f$_{max}$ values (~48 GHz) while maintaining high breakdown voltages (~192 V) [8]. Furthermore, integration of β-Ga$_2$O$_3$ onto SiC substrates have enabled unprecedented output power (2.3 W/mm at 2 GHz) and pushed f$_{max}$ up to 57 GHz [9].

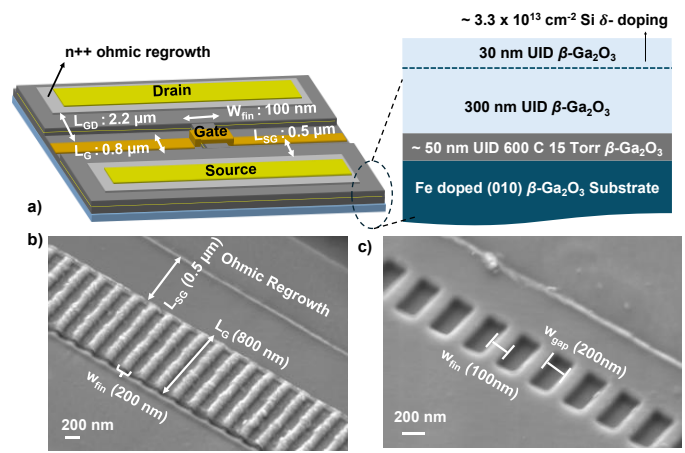

Fig 1. a) Schematic of β-Ga$_2$O$_3$ delta-doped FinFET and MOCVD grown epitaxial layer structure with delta-doped channel b) SEM Image showing the fabricated FinFET device c) SEM Image showing fins with 100 nm width and 200 nm spacing.

For high performance RF devices, it is critical to use vertically scaled channels due to the requirement of scaled gate lengths. β-(AlGa)$_2$O$_3$/Ga$_2$O$_3$ modulation doped heterostructures are effective in this respect, however these structures are currently limited to low 2DEG densities (<5x10$^{12}$ cm$^{-2}$) in β-Ga$_2$O$_3$ and are not optimal for high power density RF devices. Si delta doped channels are an attractive option since they offer the ability to tune and increase the channel charge density (n$_S$) [6], [10]. Furthermore, delta-doping, with its highly localized doping profile, enhances carrier transport properties by reducing ionized impurity scattering achieving room temperature mobilities exceeding 100 cm$^2$/V-s [10], [11], [12], [13]. Prior reports have demonstrated Si delta-doped lateral FETs with high ON-current densities, high unity current gain cut-off frequencies, and notable breakdown characteristics [14]. However, to achieve the full potential of β-Ga$_2$O$_3$, it is critical to increase the channel charge density. This enables achieving lower contact resistance, lower access region resistance and higher drain current densities. Furthermore, a Fin Field Effect Transistor design (FinFET) [15], [16], [17], [18] offers superior electrostatic control through the multi-gate, three-dimensional geometry, enabling effective modulation of high $n_s$ channels, and significantly reducing short-channel effects like drain-induced barrier lowering (DIBL), subthreshold swing degradation, and leakage currents. Additionally, the FinFET configuration could also lead to improved thermal management due to its inherently larger surface-to-volume ratio, which could be critical for β-Ga$_2$O$_3$.

In this work, we report the demonstration of MOCVD grown β-Ga$_2$O$_3$ delta-doped FETs with 100 nm fin channels modulating a record-high channel charge density of 3.3x10$^{13}$ cm$^{-2}$, along with an on/off ratio between 10$^8$-10$^9$. The work highlights the combined advantages of high quality MOCVD Si delta-doped channels with superior device geometries, to collectively push the boundaries of β-Ga$_2$O$_3$-based electronics, paving the way for enhanced performance and broader applicability in future high-power and RF electronic systems.

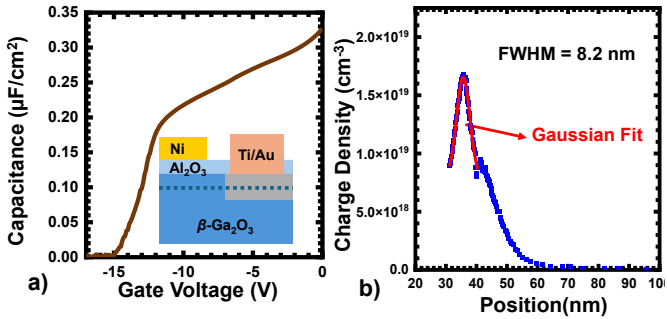

Fig 2. a) Capacitance-Voltage profile inset delta-doped planar MOSCAP b) Electron concentration profile extracted from CV measurements.

## II. DEVICE GROWTH & FABRICATION

The δ-doped epitaxial structure (Figure 1(a)) used in this work was grown by MOCVD in an Agnitron Agilis 100 system. First, ~50 nm of UID β-Ga$_2$O$_3$ was grown at 600°C, 15 Torr using Ar carrier gas and Triethylgallium (TEGa) and O$_2$ as precursors to trap the residual Fe tail [19] followed by a ~300 nm UID buffer. For the δ-doped layer, the sample was dosed with ~20 nmol of silane post a TEGa purge followed by a final cap layer of ~30 nm of UID β-Ga$_2$O$_3$. An ohmic contact regrowth process was used to achieve low contact resistance. The source and drain regions were first patterned with a thick PECVD grown SiO$_2$ hard mask. Post patterning, the β-Ga$_2$O$_3$ epilayer was etched to a depth of 60 nm and regrown with heavily doped n-type β-Ga$_2$O$_3$ with doping concentration of 1 × 10$^{20}$ cm$^{-3}$ to achieve low contact resistance [6]. Fin structures (100 nm width ($W_f$), 200 nm spacing ($W_s$)) were defined using e-beam lithography and then etched to 100 nm depth using a low power BCl$_3$ based ICP-RIE process to minimize sidewall etch damage to the δ-doped channel. A bilayer composite mask consisting of 10 nm SiO$_2$ and ZEP 520A ebeam resist was used to carry out the fin etch process. After defining the fins, 200 nm mesa etch was performed for device isolation. To reduce gate leakage, 7 nm thick Al$_2$O$_3$ was deposited via plasma assisted ALD as the gate dielectric. The Al$_2$O$_3$ over the ohmic regions was selectively removed using BCl$_3$ etching for source/drain metallization. A Ti/Au (40nm/60nm) stack was then deposited via e-beam evaporation followed by short RTA anneal (470°C, 1 min, N$_2$ ambient) to form the ohmic contacts. Finally, a 100 nm Schottky Ni gate was deposited using e-beam evaporation with 180-degree sample rotation for even deposition on the gate sidewall. The fabricated device is imaged using scanning electron microscope and shown in Figure 1(b) and 1(c).

## III. RESULTS & DISCUSSION

Capacitance-voltage (CV) measurements were carried out using an Agilent B1500 parameter analyzer at a frequency of 100 kHz (Figure 2(a)) on planar MOS structures (see inset of Figure 2(a)). The carrier density profile extracted (Figure 2(b)) shows a narrow doping distribution with an integrated sheet charge density of ~2.8×10$^{13}$ cm$^{-2}$ and full width half maximum (FWHM) of 8.2 nm. Hall measurements revealed a δ-doped sheet carrier density of ~3.3 x 10$^{13}$ cm$^{-2}$ with Hall mobility of 89 cm$^2$/Vs at room temperature. The discrepancy between CV and Hall may be attributed partly to the high leakage in the MOS structures for voltages exceeding -10 V. Contact Resistance of 0.42 Ω-mm and a sheet resistance of 2.04 kΩ/□ were measured using transfer length measurements (TLM) (Figure 3(a)). The sheet resistance extracted from TLM measurements matches closely to that estimated using Hall (2.1 kΩ/□).

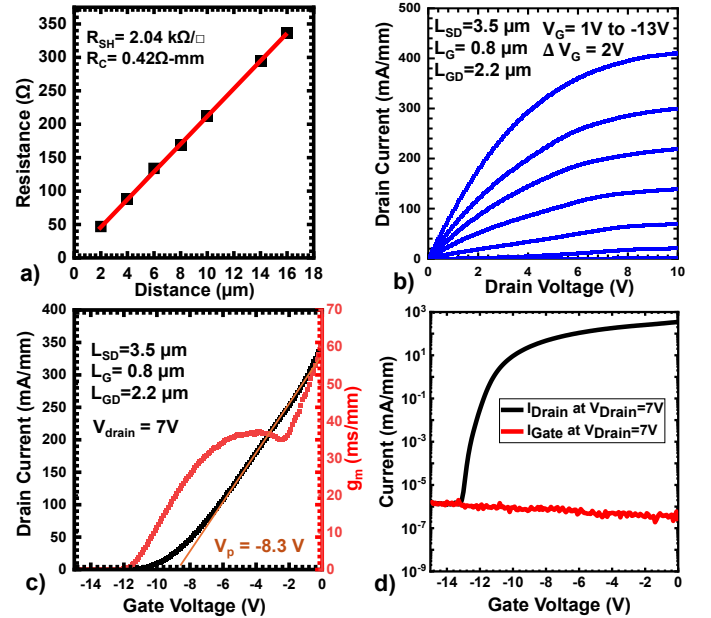

Fig 3. a) TLM measurements showing sheet resistance and total contact resistance. b) Output IV characteristics, c) transfer IV characteristics and d) Log-scale transfer IV characteristics of delta-doped FinFET.

Figure 3(b) shows the representative DC output characteristics of the delta doped FinFET with a gate length of 0.8 μm. A maximum drain current of 410 mA/mm was measured at V$_{GS}$=1V and V$_{DS}$=10 V, which is the highest reported for β-Ga$_2$O$_3$ FinFETs. The drain current is normalized to the effective device width which in this case is 33% of the total width (W$_f$=100 nm, W$_s$ =200 nm). The on-resistance of the device is measured to be 10.1 Ω.mm. A maximum transconductance (g$_{m,max}$) of 60 mS/mm was measured at a gate bias of 0 V (Figure 3(c)). The transconductance shows an increase after an initial peak at -4V which suggests the possibility of fixed interface charges at the $Al_2O_3/Ga_2O_3$ interface leading to the formation of an accumulation channel. A pinch off voltage of ~ -8.3V was estimated by linear extrapolation of transfer I-V which is much lower than that obtained on planar MOS structures (~ -15 V, inset of Figure 3(a)). The lowered $V_P$ is attributed to the sidewall depletion in the case of FinFET which is absent in the planar devices. The measured FinFETs also display a high $I_{ON}/I_{OFF}$ ratio between $10^8$ -$10^9$ even with $n_s$> $3\times10^{13}$ cm$^{-2}$ (Figure 3(d)), showing the effectiveness of the FinFET design.

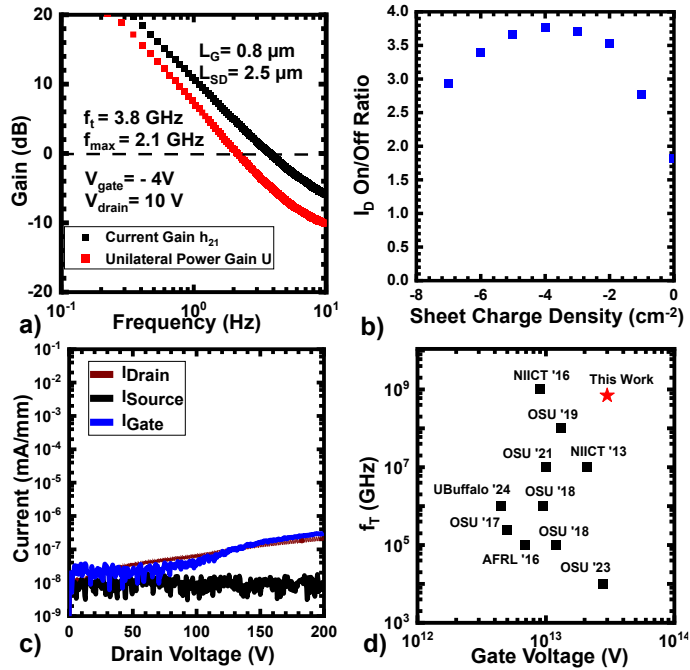

Fig. 4 a) Measured small signal characteristics of 0.8 µm gate length delta-doped FinFET b) Measured cutoff frequency $f_T$ as a function of gate voltage measured at 10 V drain bias c) Three terminal breakdown characteristics showing breakdown voltage >200 V d) Benchmarking plot comparing $n_s$ and $I_{ON}/I_{OFF}$ ratio of β-$Ga_2O_3$ FETs.

The high frequency small-signal performance was characterized between 0.1 GHz and 25 GHz using an vector network analyzer. Figure 4(a) shows the measured post de-embedded short-circuit current gain (h21) & unilateral power gain (U). Peak current gain cut-off frequency ($f_T$) of 3.8 GHz and associated peak maximum oscillation frequency ($f_{MAX}$) of 2.1 GHz were measured at gate bias of -4 V and drain bias of 10 V. The lower $f_{MAX}$ is likely due to large gate resistance due from the use of thin (100 nm) Ni gates, which is more resistive compared to Au. The dependence of the cutoff frequency on gate bias (shown in Figure 4(b)) follows a similar trend to that of the $g_m$, peaking at -4 V, however $f_T$ does not follow the increase in $g_m$ observed above -3 V. The three terminal breakdown performance of the FinFETs were measured till 200 V (see Figure 4(c)). The leakage current from all the three terminals remained well below nominal breakdown condition (0.1 mA/mm) demonstrating high voltage performance of these devices even with $n_s$> $3\times10^{13}$ cm$^{-2}$. Figure 4(d) compares the $I_{On}/I_{Off}$ ratio and sheet charge density ($n_s$) of various reported β-$Ga_2O_3$-based field-effect transistors (FETs), serving as a benchmark for evaluating the performance of our device [20], [21], [22], [23]. Positioned toward the upper right region of the plot, our device demonstrates a competitive $I_{On}/I_{Off}$ ratio indicating great FET operation along with high channel charge modulation. This performance underscores the effectiveness of our design and fabrication strategy in pushing the limits of power device operation in the ultra-wide bandgap regime.

## IV. CONCLUSION

In conclusion, we have demonstrated MOCVD-grown β-$Ga_2O_3$ Si delta-doped FinFETs, modulating a record high channel charge density of $3.3\times10^{13}$ cm$^{-2}$ along with a high on/off ratio of ~$10^8$-$10^9$. Using a 100 nm fin width and a high charge density Si delta-doped channel, the devices exhibit record high performance including drain current density of 410 mA/mm, transconductance of 60 mS/mm, breakdown voltage exceeding 200V and current gain cut off frequency of 3.8 GHz. The DC and RF performance of these FinFETs can be further enhanced by scaling the gate length.

## REFERENCES

[1] A. Kumar, K. Ghosh, and U. Singisetti, "Low field transport calculation of 2-dimensional electron gas in β-$(Al_xGa_{1−x})_2O_3/Ga_2O_3$ heterostructures," *J. Appl. Phys.*, vol. 128, no. 10, Sep. 2020, doi: 10.1063/5.0008578.

[2] "Low-field and high-field transport in β-Ga2O3," in *Gallium Oxide*, Elsevier, 2019, pp. 149–168. doi: 10.1016/b978-0-12-814521-0.00007-5.

[3] Z. Xie, C.-K. Tan, and M.-C. Cheng, "High-field electron transport properties of β-Ga2O3: An integrated Monte Carlo and first-principles approach," *Appl. Phys. Lett.*, vol. 126, no. 21, May 2025, doi: 10.1063/5.0271859.

[4] E. Johnson, "Physical limitations on frequency and power parameters of transistors," in *IRE International Convention Record*, New York, NY, USA: Institute of Electrical and Electronics Engineers, pp. 27–34. doi: 10.1109/irecon.1965.1147520.

[5] A. J. Green *et al.*, "$\beta$ -Ga2O3 MOSFETs for Radio Frequency Operation," *IEEE Electron Device Lett.*, vol. 38, no. 6, pp. 790–793, Jun. 2017, doi: 10.1109/LED.2017.2694805.

[6] Z. Xia *et al.*, "Delta Doped $\beta$ -Ga2O3 Field Effect Transistors With Regrown Ohmic Contacts," *IEEE Electron Device Lett.*, vol. 39, no. 4, pp. 568–571, Apr. 2018, doi: 10.1109/LED.2018.2805785.

[7] A. Vaidya, C. N. Saha, and U. Singisetti, "Enhancement Mode β-$(Al_xGa_{1-x})_2O_3/Ga_2O_3$ Heterostructure FET (HFET) With High Transconductance and Cutoff Frequency," *IEEE Electron Device Lett.*, vol. 42, no. 10, pp. 1444–1447, Oct. 2021, doi: 10.1109/led.2021.3104256.

[8] C. N. Saha *et al.*, "Scaled β-Ga2O3 thin channel MOSFET with 5.4 MV/cm average breakdown field and near 50 GHz fMAX," *Appl. Phys. Lett.*, vol. 122, no. 18, May 2023, doi: 10.1063/5.0149062.

[9] M. Zhou *et al.*, "1.1 A/mm ß-Ga$_2$O$_3$-on-SiC RF MOSFETs with 2.3 W/mm P$_{out}$ and 30% PAE at 2 GHz and f$_T$/f$_{max}$ of 27.6/57 GHz," in *2023 International Electron Devices Meeting (IEDM)*, San Francisco, CA, USA: IEEE, Dec. 2023, pp. 1–4. doi: 10.1109/iedm45741.2023.10413782.

[10] S. Krishnamoorthy, Z. Xia, S. Bajaj, M. Brenner, and S. Rajan, "Delta-doped β-gallium oxide field-effect transistor," *Appl. Phys. Express*, vol. 10, no. 5, p. 051102, May 2017, doi: 10.7567/APEX.10.051102.

[11] C. Joishi *et al.*, "Deep-Recessed $\beta$-Ga$_2$O$_3$ Delta-Doped Field-Effect Transistors With *In Situ* Epitaxial Passivation," *IEEE Trans. Electron Devices*, vol. 67, no. 11, pp. 4813–4819, Nov. 2020, doi: 10.1109/TED.2020.3023679.

[12] N. Kumar, C. Joishi, Z. Xia, S. Rajan, and S. Kumar, "Electrothermal Characteristics of Delta-Doped $\beta$-Ga$_2$O$_3$ Metal–Semiconductor Field-Effect Transistors," *IEEE Trans. Electron Devices*, vol. 66, no. 12, pp. 5360–5366, Dec. 2019, doi: 10.1109/TED.2019.2944628.

[13] P. Ranga *et al.*, "Delta-doped $\beta$-Ga$_2$O$_3$ thin films and $\beta$-(Al$_{0.26}$Ga$_{0.74}$)$_2$O$_3$/$\beta$-Ga$_2$O$_3$ heterostructures grown by metalorganic vapor-phase epitaxy," *Appl. Phys. Express*, vol. 13, no. 4, p. 045501, Apr. 2020, doi: 10.35848/1882-0786/ab7712.

[14] Z. Xia *et al.*, "$\beta$-Ga$_2$O$_3$ Delta-Doped Field-Effect Transistors With Current Gain Cutoff Frequency of 27 GHz," *IEEE Electron Device Lett.*, vol. 40, no. 7, pp. 1052–1055, Jul. 2019, doi: 10.1109/LED.2019.2920366.

[15] G. Amarnath, D. Krishna, and A. Vinod, "TCAD-based Comparative Study of Gallium-Oxide based FinFET and MOSFET," in *2020 IEEE International Conference on Advent Trends in Multidisciplinary Research and Innovation (ICATMRI)*, Buldhana, India: IEEE, Dec. 2020, pp. 1–4. doi: 10.1109/ICATMRI51801.2020.9398440.

[16] K. D. Chabak *et al.*, "Enhancement-mode Ga2O3 wrap-gate fin field-effect transistors on native (100) $\beta$-Ga2O3 substrate with high breakdown voltage," *Appl. Phys. Lett.*, vol. 109, no. 21, p. 213501, Nov. 2016, doi: 10.1063/1.4967931.

[17] H. Ebrahimi-Darkhaneh, L. Fernandez-Izquierdo, J. Arellano-Jimenez, M. Quevedo-Lopez, and S. K. Banerjee, "Fabrication and Characterization of Ga$_2$O$_3$ FinFETs on Patterned Silicon Substrate," *Adv. Electron. Mater.*, p. 2400945, Apr. 2025, doi: 10.1002/aelm.202400945.

[18] Y. Wei *et al.*, "Low Reverse Conduction Loss β-Ga$_2$O$_3$ Vertical FinFET With an Integrated Fin Diode," *IEEE Trans. Electron Devices*, vol. 70, no. 7, pp. 3454–3461, Jul. 2023, doi: 10.1109/TED.2023.3274499.

[19] J. P. McCandless *et al.*, "Accumulation and removal of Si impurities on $\beta$-Ga2O3 arising from ambient air exposure," *Appl. Phys. Lett.*, vol. 124, no. 11, p. 111601, Mar. 2024, doi: 10.1063/5.0191280.

[20] N. K. Kalarickal *et al.*, "Demonstration of self-aligned β-Ga2O3 δ-doped MOSFETs with current density >550 mA/mm," *Appl. Phys. Lett.*, vol. 122, no. 11, Mar. 2023, doi: 10.1063/5.0131996.

[21] M. H. Wong, K. Sasaki, A. Kuramata, S. Yamakoshi, and M. Higashiwaki, "Field-Plated Ga$_2$O$_3$ MOSFETs With a Breakdown Voltage of Over 750 V," *IEEE Electron Device Lett.*, vol. 37, no. 2, pp. 212–215, Feb. 2016, doi: 10.1109/LED.2015.2512279.

[22] H. Murakami *et al.*, "Homoepitaxial growth of β-Ga$_2$O$_3$ layers by halide vapor phase epitaxy," *Appl. Phys. Express*, vol. 8, no. 1, p. 015503, Jan. 2015, doi: 10.7567/apex.8.015503.

[23] M. Higashiwaki *et al.*, "Depletion-mode Ga2O3 metal-oxide-semiconductor field-effect transistors on β-Ga2O3 (010) substrates and temperature dependence of their device characteristics," *Appl. Phys. Lett.*, vol. 103, no. 12, Sep. 2013, doi: 10.1063/1.4821858.